\documentclass[aps,prl,onecolumn,groupedaddress,amsmath,amssymb]{revtex4-1} 
\usepackage{upgreek}
\usepackage{graphicx}  
\usepackage{dcolumn}   
\usepackage{bm}        
\usepackage{verbatim}   
\usepackage[sort&compress]{natbib} 		   

\expandafter\def\expandafter\normalsize\expandafter{%
	\normalsize
	\setlength\abovedisplayskip{15pt}
	\setlength\belowdisplayskip{15pt}
	\setlength\abovedisplayshortskip{15pt}
	\setlength\belowdisplayshortskip{15pt}
}                                                                  
                                                                  
\begin{document}

\title{Jarzynski Equality for Soret Equilibria -- Virtues of Virtual Potentials} 
\author{Tobias Thalheim}
\affiliation{Molecular Nanophotonics Group, Peter Debye Institute for Soft Matter Physics, Leipzig University, 04103 Leipzig, Germany}
\author{Marco Braun}
\affiliation{Molecular Nanophotonics Group, Peter Debye Institute for Soft Matter Physics, Leipzig University, 04103 Leipzig, Germany}
\author{Gianmaria Falasco}
\affiliation{Theory of Condensed Matter, Institute for Theoretical Physics, Leipzig University, Postfach 100 920, 04009 Leipzig, Germany}
\affiliation{Complex Systems and Statistical Mechanics, Department of Physics and Materials Science, University of Luxembourg, L-1511, Luxembourg}
\author{Klaus Kroy}
\affiliation{Theory of Condensed Matter, Institute for Theoretical Physics, Leipzig University, Postfach 100 920, 04009 Leipzig, Germany}
\author{Frank Cichos}
\affiliation{Molecular Nanophotonics Group, Peter Debye Institute for Soft Matter Physics, Leipzig University, 04103 Leipzig, Germany}
   
\date{\today}
\begin{abstract}
The Jarzynski equality relates the free energy difference between two equilibrium states to the fluctuating irreversible work afforded to switch between them. The prescribed fixed temperature for the equilibrium states implicitly constrains the dissipative switching process that can take the system far from equilibrium. Here, we demonstrate theoretically and experimentally that such a relation also holds for the nonisothermal case, where the initial stationary state is not in equilibrium and the switching is effected by dynamically changing temperature gradients instead of a conservative force.  Our demonstration employs a single colloidal particle trapped by optically induced thermophoretic drift currents. It relies on identifying suitable equivalents of classical work and heat and our ability to measure their distributions and express them in terms of a virtual potential. 
\newline
\end{abstract}

\maketitle
\section{I. Introduction}
Virtually all interesting processes in nature and technology down to the smallest length scales are irreversible non-equilibrium processes \cite{Helgeson:GeoCosActa:1968, Margretier:JBioenBiomem:1995, Agard:BioChem:1994,Lior:Energy:2003, Guenter:Macromol:2003, Saalwaechter:Macromol:2013}, proceeding even far from equilibrium\cite{Turlier:NatPhys:2016,Bechinger:RevModPhys:2016, Falasco:PhysRevE:2016} and, therefore, cannot be treated by classical thermodynamics \cite{Argun:PhysRevE:2016}. In those non-equilibrium cases, fluctuation theorems still provide symmetry relations to describe the system's evolution in terms of entropy production \cite{Morriss:PhysRevLett:1993, Cohen:JStatPhys:1995, Kurchan:JPhysAMathGen:1998, Spohn:JStatPhys:1999}. In this group of fluctuation theorems, particularly the Crooks relation can be used to assess the degree of irreversibility of microscopically reversible non-equilibrium processes by relating the work probability distribution along the forward and reversed processes \cite{Crooks:JStatPhys:1998, Crooks:PhysRevE:1999}:
\begin{equation}\label{eq:CrooksRelation}
\frac{\mathcal{P}_{\textrm{forward}}(W)}{\mathcal{P}_{\textrm{backward}}(-W)} = e^{-\beta(W-\Delta F)}
\end{equation}
with $\beta = (k_{\mathrm{B}}T_0)^{-1}$, $k_{\mathrm{B}}$ the Boltzmann constant and $T_0$ the temperature of the heat bath coupled to the perturbed system. In this equation, $\Delta F$ is the free energy difference of the system's initial state A and its final state B. $\mathcal{P}_{\textrm{forward}}$ ($\mathcal{P}_{\textrm{backward}}$) is the probability of the work $W$ ($-W$) over all paths $\gamma$ that yield the value $W$ ($-W$). In other words, Equation (\ref{eq:CrooksRelation}) relates the probability of dissipating a default amount of work along $\gamma$ to the probability of extracting the identical amount of work from the heat bath at $T_{0}$ in the time-reversed process which is smaller than the probability of the forward process \cite{Gomez:IntJThermodyn:2013}. Note that the system starts from a thermal equilibrium state A, but the final state B is not required to be in equilibrium immediately after the perturbation ceases \cite{Bustamante:Nature:2005}. 

A non-equilibrium equality can be derived from Crooks fluctuation theorem \cite{Crooks:JStatPhys:1998, Crooks:PhysRevE:1999} which allows to determine the free energy difference $\Delta F$ between the two states A and B with the help of the work $W$ applied -- the Jarzynski equality (JE) \cite{Jarzynski:PhysRevLett:1997}:
\begin{equation}\label{eq:JarzynskiEquality}
\left \langle e^{-\beta W} \right \rangle=e^{-\beta \Delta F}.
\end{equation}
The brackets $\left \langle \cdot \right \rangle$ denote the average over an infinite number of experiments with the same initial and final states A and B, but with any path $\gamma$ connecting them. The strength of the JE is that it is exact for calculating $\Delta F$ between the two states A and B for a given work, whereas the second law of thermodynamics in terms of work and free energy differences, which can be derived from the JE (cf. for example \cite{Gomez:IntJThermodyn:2013}), merely provides an upper bound for the free energy difference: $\left \langle W \right \rangle \geq \Delta F$. In this situation, the equality sign solely holds for reversible processes, where all states along $\gamma$ between A and B are still in their thermodynamic equilibria demanding an infinitely slow perturbation process on microscopic length scales. 

Meanwhile, the JE has been validated for a multitude of experiments from different research topics \cite{Liphardt:Science:2001, Liphardt:Science:2002, Wang:PhysRevLett:2002, Collin:Nature:2005, Huber:PhysRevLett:2008, An:NatPhys:2015} or even in macroscopic systems \cite{Douarche:EPL:2005}. The verification of the JE in a time-dependent non-harmonic potential was done by Bechinger et al. \cite{Blicke:PhysRevLett:2006}. Common to all these examples is that the time-dependent protocol $\lambda = \lambda (t)$ for driving the corresponding system out of thermal equilibrium always generates an external perturbation force with the help of real potentials, i.e., work is transferred from the exterior to bring the system of interest from the initial state A to the final state B. As an extension to real potentials, Bechhoefer et al. \cite{Bechhoefer:PhysRevE:2012, Bechhoefer:PhysRevLett:2014} implemented a version of the anti-Brownian electrokinetic trap \cite{Moerner:ApplPhysLett:2005, Cohen:Phd:2006} to generate a virtual potential utilizing feedback loops and confirmed the JE as well. Further theoretical research unveiled that under certain assumptions on the system's steady state distribution function, the requirement of the states A and B to be in thermodynamic equilibrium can be dropped and the JE still holds \cite{Hatano:PhysRevE:1999}. 

\begin{figure}
\includegraphics[scale=0.25]{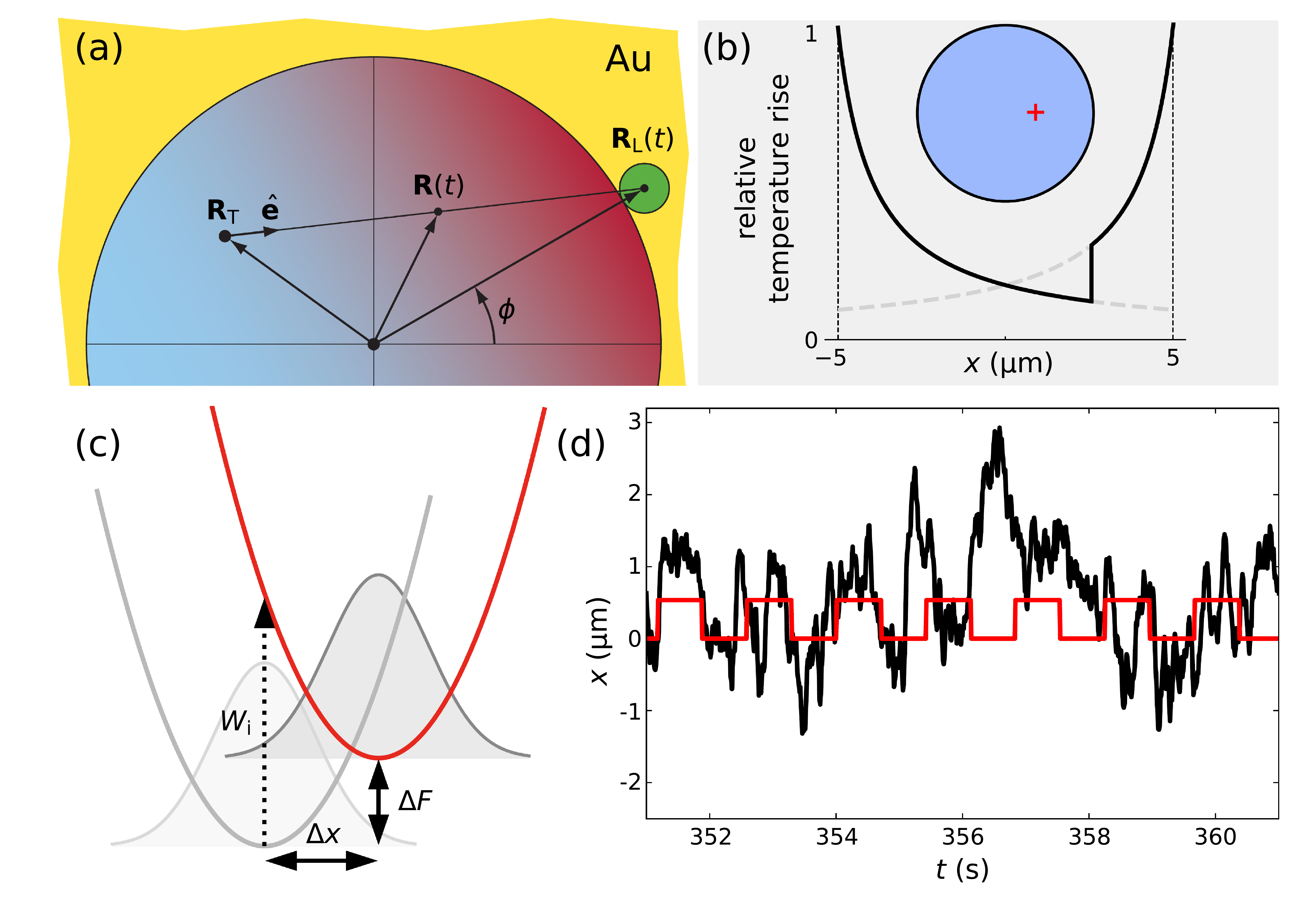}
\caption{\label{Figure1} (a) Dynamic temperature gradients are generated by focusing a laser beam on the circumference of a circular gold structure ($\bm R_\mathrm{L}(t)$). A feedback algorithm detects a particle ($\bm R(t)$) in the circular trap in real time and re-positions the laser to push the particle towards a target in the trap ($\bm R_\mathrm{T}$) via thermophoresis. (b) Simulated temperature profile of the feedback trap. The red cross indicates the trapping target. (c) Switching the virtual trapping potential between the initial steady state A to the final steady state B separated by $\Delta x$. The gray-colored bell-shaped plots represent the particle's probabilities of presence in the corresponding states. (d) Trajectory of a particle (black) during the switching of the virtual trapping potential (red).}
\end{figure}

We test the JE for a single colloid trapped by dynamic temperature fields in a thermophoretic trap as introduced before \cite{Braun:PhysChemChemPhys:2014, Braun:NanoLett:2015, Fraenzl:NatMethods:2019, Baffou:NatMater:2020} (cf. Fig. \ref{Figure1}(a), (b)). The trapping of the colloid is the result of a thermodynamical non-equilibrium process known as Soret effect or thermophoresis which is driving the colloid to colder regions as a result of interfacial flows. In our realization, a steady state is achieved by a feedback process quickly switching the directions of a temperature gradient. 
The local temperature minimum trapping the colloid is, consequently, only virtual and existing solely in the time average. Further, no conservative force is driving the particle \cite{Braun:ACSNano:2013} as compared to all previous tests. To explore the Jarzynski equality, the position of this virtual temperature minimum is actively toggled between the two steady state positions according to 
\begin{equation}\label{eq:JumpingTemperatureField}
\Delta T (x)=\frac{1}{2}\alpha \left[x-\lambda(t)\right]^2
\end{equation}
as shown in Fig. \ref{Figure1}(c). $\alpha$ describes the curvature of the temperature field. The time-dependent protocol $\lambda(t)$ is a square wave step function accounting for a periodic switching of the harmonic virtual temperature field between two central points at a switching time $\tau$ as shown in Fig. \ref{Figure1}(d) in red \cite{ Bechhoefer:PhysRevLett:2014}. Dissipative thermo-osmotic processes driving the thermophoresis of the colloid \cite{Bregulla:PhysRevLett:2016} are stronger in the regions with higher temperature gradients and force the colloid to relax to the minimum of the corresponding temperature profile. Note that no external potential (or resulting conservative force) causes this motion. The equivalent quantities for the work and heat distribution in the thermal non-equilibrium are measured and calculated. The results show that an equivalent formulation of the JE exists which involves gradients of temperature fields and not potentials of conservative forces. Moreover, the initial and final states A and B are not required to be in equilibrium anymore. They correspond to steady states of the non-equilibrium effect thermophoresis (Soret ``equilibrium'', cf. \cite{Wuerger:ComptesRendusMecanique:2013}). 

\section{II. Soret ``Equilibrium''}
The physical principle behind the thermophoretic trap is a non-equilibrium process called thermophoresis. Colloidal thermophoresis refers to the motion of suspended particles due to a temperature gradient $\nabla T(\boldsymbol{r})$ within the suspension liquid. Temperature gradients along the colloid--solvent interface generate an interfacial tension in the liquid  parallel to $\nabla T(\boldsymbol{r})$ due to the flow of heat and the excess interaction of colloid and solvent \cite{Piazza:SoftMatter:2008, Prost:EurPhysJE:2009, Wuerger:RepProgPhys:2010}. This interfacial tension causes the liquid to flow from cold to hot regions in a thin layer at the interface \cite{Bregulla:PhysRevLett:2016}. This thermo-osmotic flow is balanced by a motion of the particle from hot to cold regions. It, therefore, relies on a force balance and no net body forces on the particle are present in this case making it force-free. Yet, the interfacial flows continuously dissipate energy during the motion of the particle. 

According to Onsager's relations \cite{Wuerger:ComptesRendusMecanique:2013}, this motion occurs in addition to Brownian motion, consequently, the total probability density flux can be written as
\begin{equation}\label{eq:FluxDensity}
\boldsymbol{J} = -D\nabla P - PD_{\mathrm{T}}\nabla T,
\end{equation}
where $D$ is the Brownian diffusion coefficient, which is assumed to be approximately constant within the trapping region. The term $P = P(\boldsymbol{r})$ describes the probability density distribution for finding a colloid at a particular position $\boldsymbol{r}$ in the inhomogeneous temperature landscape $T(\boldsymbol{r}) = T_{0} + \Delta T(\boldsymbol{r})$. The local temperature rise $\Delta T(\boldsymbol{r})$ is typically small compared to the ambient temperature $T_{0} = 296~\mathrm{K}$ for our experiments \cite{Braun:ACSNano:2013, Parola:JPhysCondMat:2008}. We, therefore, assume the thermophoretic mobility $D_{\mathrm{T}}$ to be constant within our thermophoretic trap. The steady state probability density distribution (Soret ``equilibrium'') can be obtained by balancing the diffusive and thermo-diffusive parts in Equation (\ref{eq:FluxDensity}) (i.e., $\boldsymbol{J} = 0$) yielding an exponential dependence for $P(\boldsymbol{r})$:
\begin{equation}\label{eq:SoretEquilibrium}
\frac{P}{P_0}=\exp\left(-S_{\rm T} \Delta T\right).
\end{equation}
The parameter $S_{\mathrm{T}} = \frac{D_{\mathrm{T}}}{D}$ is called Soret coefficient and is positive for a motion in the opposite direction of the temperature gradient. Values in the range of $S_{\mathrm{T}}=$ 0.01-10$\,\rm K^{-1}$ are typically found. The specific value of $S_{\mathrm{T}}$ is depending on a variety of parameters as, for example, temperature \cite{Piazza:EPL:2003, Wong:Langmuir:2007}, pH value of the solvent \cite{Piazza:EPL:2003, Braun:PNAS:2006} or particle size \cite{Braun:PhysRevLett:2006, Piazza:PhysRevLett:2008}. A virtual effective potential energy landscape can be defined from Equation (\ref{eq:SoretEquilibrium}) by comparing to a Boltzmann distribution which results in \cite{Duhr:PhysRevLett:2006, Braun:NanoLett:2015}
\begin{equation}\label{eq:EquivalentPotentialDefinition}
\frac{U_{\rm eff}}{k_{\rm B}T} \sim S_{\rm T} \Delta T.
\end{equation} 
The potential $U_{\rm eff}$ hereby corresponds to the potential energy landscape that would have to be created by an external force to achieve the same confinement as due to the action of the temperature field. In combination with Equation (\ref{eq:JumpingTemperatureField}), the virtual effective potential energy can be written as
\begin{equation}\label{eq:EquivalentPotential}
U_{\rm eff} (x)=\frac{1}{2}\kappa_{\rm eff} \left[x-\lambda(t)\right]^2
\end{equation}
with an effective trapping stiffness $\kappa_{\rm eff} = k_\mathrm{B}TS_\mathrm{T}\alpha$. Due to the feedback control of our thermophoretic trap the effective potential energy landscape is virtual and can be exactly adjusted, which is used in the subsequent calibration procedure.

\section{III. Calibration of the thermophoretic trap}
\begin{figure}
\includegraphics[scale=0.35]{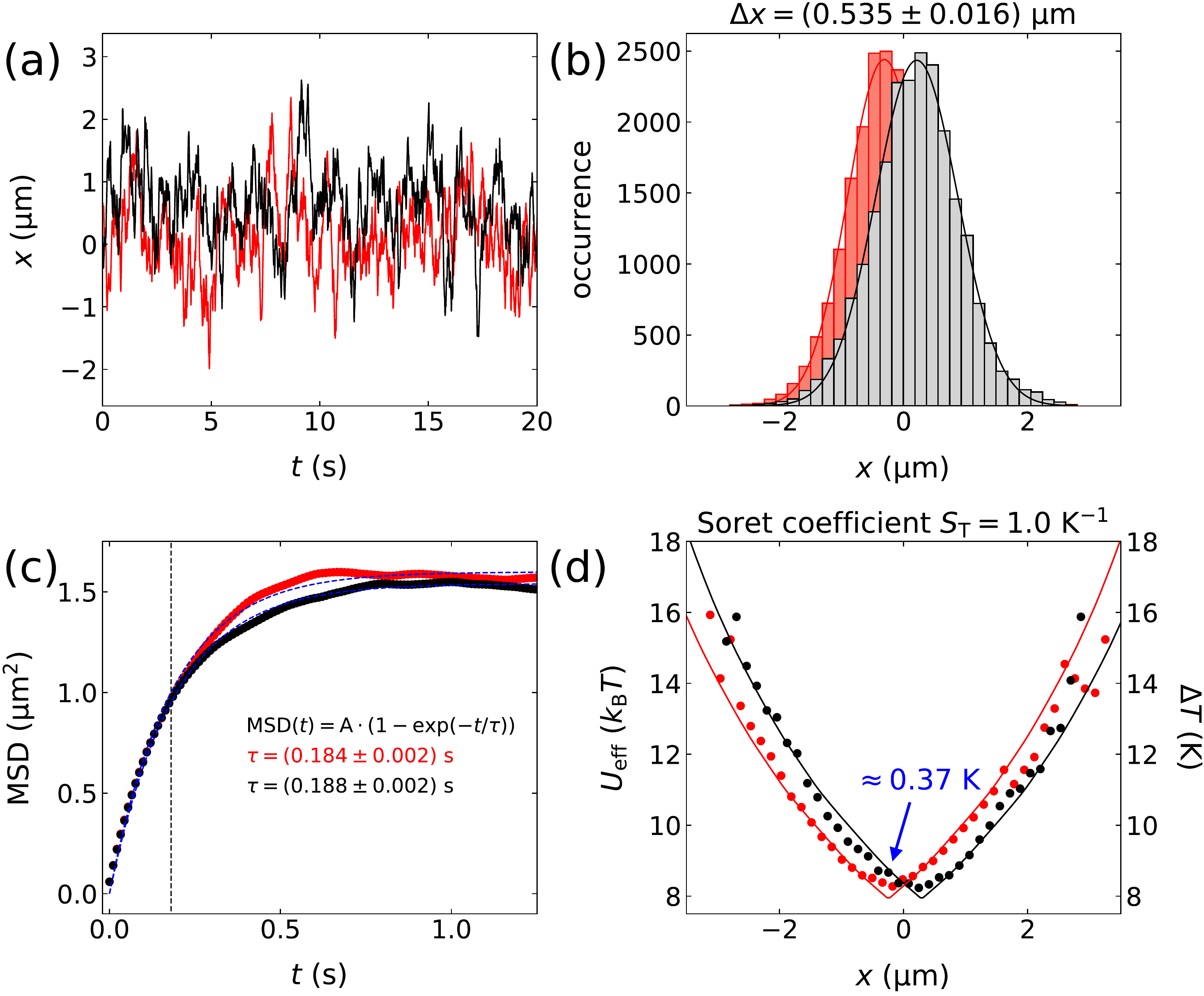}
\caption{\label{Figure2} (a) Trajectories of the 200 nm PS particle in each virtual effective potential. (b) Steady-state probability distribution obtained from the particle trajectories. (c) MSD calculated from the trajectories of each virtual effective potential. The dashed blue lines are fits to the data according to a Brownian particle in a harmonic potential \cite{Doi:Book:1986}. The vertical dashed black line shows the intrinsic relaxation time of the MSD. (d) Temperature distribution for each potential obtained from experiments (data points) and from simulations (lines) \cite{Braun:NanoLett:2015}. The average temperature change $\langle \Delta T_\mathrm{W} \rangle = \langle \Delta T_\mathrm{Q} \rangle \approx 0.37~\mathrm{K}$ driving the particle between the two targets can be seen in blue.} 
\end{figure}

Our thermophoretic trap is used in feedback mode \cite{Braun:NanoLett:2015}, i.e., the heating laser is placed dynamically on the gold structure to yield a temperature gradient that pushes the particle towards a target position. On the relevant length scales heat diffusion is much faster than mass diffusion and, thus, temperature changes can be treated as instantaneous. The thermophoretic trap is calibrated for the two target positions, which are slightly displaced in space. Trajectories of a 200 nm polystyrene (PS) bead (ThermoFisher, order number: F8810) are acquired for both targets individually (see Fig. \ref{Figure2}(a)). The separation of the targets is then calculated by fitting the steady-state probability density to a Gaussian distribution (see Fig. \ref{Figure2}(b)). The distance of the maxima is the separation of both targets, which is for this experiment $\Delta x=(0.54 \pm 0.02)\,\rm \upmu m$. The widths that are related to the effective trapping stiffness via $\kappa_{\rm eff}=k_{\rm B}T/\sigma^2$ are measured to $\sigma=(0.62\pm0.02)\,\rm \upmu m$ and, consequently, $\kappa_{\rm eff}=(10.5\pm0.7)\,\rm{fN\upmu m^{-1}}$. An intrinsic relaxation time of the trapped particle is measured from the mean squared displacement (MSD) \cite{Jacobson:ARBBS:1997} and is in the order of $\tau_{\rm MSD}\approx 0.2\,\rm s$ (cf. Fig. \ref{Figure2}(c)). 

The simulated temperature profile \cite{Braun:NanoLett:2015} and $\Delta T_{\rm Au}/P_{\rm heat}\approx 29\,\rm K\,\rm mW^{-1}$, where $\Delta T_\mathrm{Au}$ is the temperature elevation on the heated gold surface for a given power $\Delta P_\mathrm{heat}$ of the focused trapping laser (cf. Supplemental Material), yield the temperature distribution shown in Fig. \ref{Figure2}(d). Comparing this temperature landscape to the measured positional distributions via Equation (\ref{eq:SoretEquilibrium}) unveils a Soret coefficient of $S_{\rm T}=1.0\,\rm K^{-1}$ for the 200 nm PS particles used in these experiments. The average temperature rise upon switching between the two target positions is $\langle \Delta T_\mathrm{W} \rangle = \langle \Delta T_\mathrm{Q} \rangle \approx 0.37~\mathrm{K}$ and, hence, very small, even though the overall temperature increment in the center of the trap is about 8 K with respect to room temperature.  

\section{IV. Test of JE for virtual effective potentials}
In order to verify the JE, we trap a single 200 nm PS bead in a virtual effective potential $U_{\rm eff}$ according to Equation (\ref{eq:JumpingTemperatureField}). The equivalent free energy $F$ of the particle in the virtual effective potential is invariant under translation of $U_\mathrm{eff}$ yielding that the "free energy" difference is zero in this case, therefore, simplifying the JE to $\left \langle e^{-\beta W} \right \rangle=1$. The switching period $\tau$ is varied to study the relaxation behavior of the particle after each switching. 

The equivalent work that is done on the particle by switching the virtual effective harmonic potential is measured from the trajectory $x(t)$ via the path integral \cite{Blicke:PhysRevLett:2006}
\begin{equation}\label{eq:AppliedWork}
W\left[x(t)\right]=\int_0^{\tau}\mathrm dt\, \frac{\partial U_{\rm eff}(x(t),t)}{\partial \lambda}\dot{\lambda}(t),
\end{equation}
which simplifies to the integration over a delta function as $\lambda$ is a square wave step function. This work is related to the energy that is needed to change the interfacial flow fields around the colloid, while additional energy is also required to maintain the temperature gradients. 

Accordingly, the dissipated heat of the particle due to the relaxation into the new potential is measured by
\begin{equation}\label{eq:DissipatedHeat}
Q\left[x(t)\right]=\int_0^{\tau}\mathrm dt\, \frac{\partial U_{\rm eff}(x(t),t)}{\partial x}\dot{x}(t).
\end{equation}
On average and for $\Delta F=0$, the work is fully dissipated to heat: $\left\langle W\right\rangle=\left\langle Q\right\rangle$.

\begin{figure}
\includegraphics[scale=0.35]{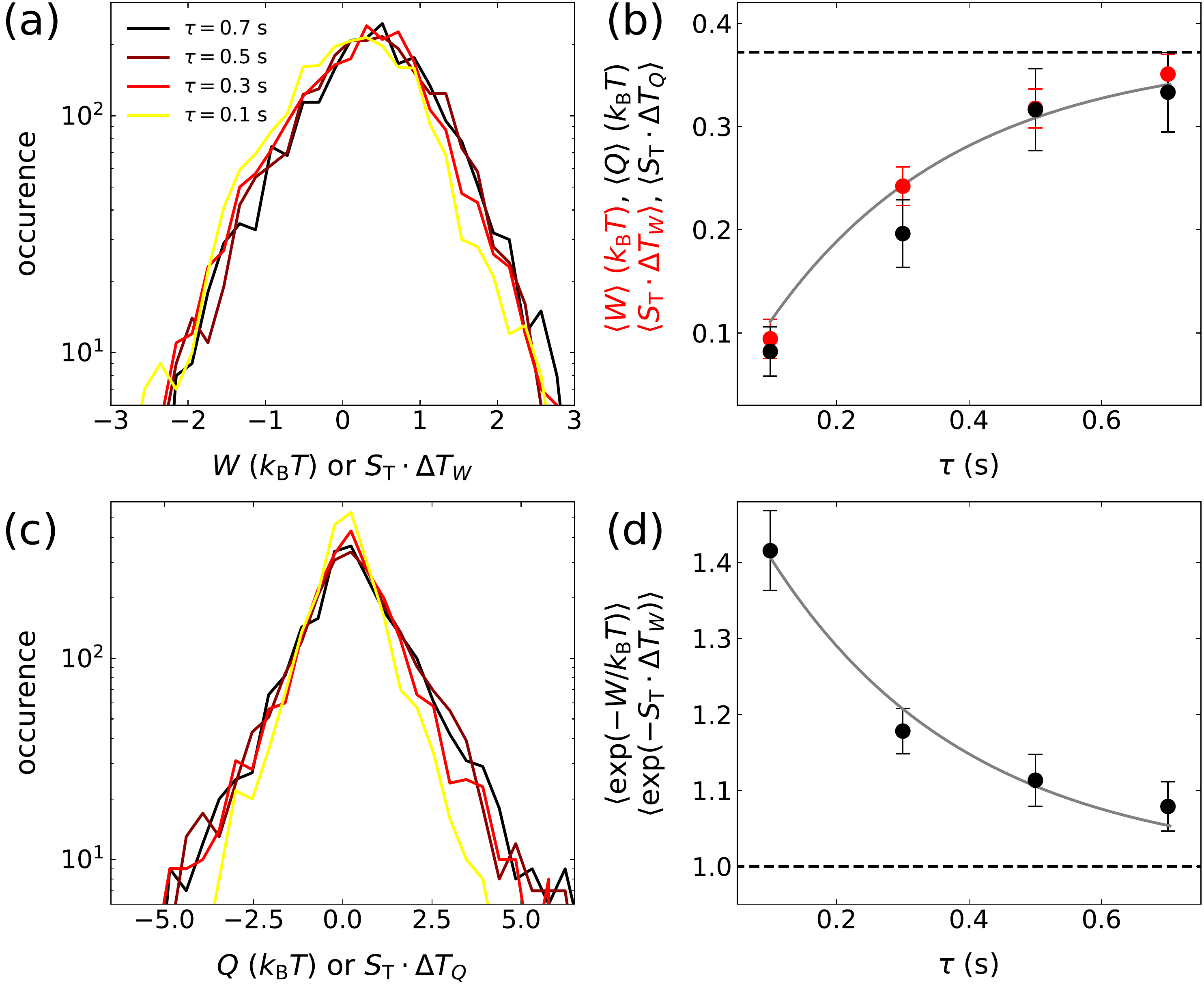}
\caption{\label{Figure3} (a) Histogram of the measured equivalent work $W$ for different switching times $\tau$. (b) Offset of the mean equivalent work (red) as well as average equivalent energy dissipated (black) in dependence on the switching times adjusted. The dashed line represents the offset for infinite large switching times. (c) Histogram of the measured heat $Q$ for different switching times $\tau$ (same color code as in (a)). (d) Exponential average over the work done (cf. Equation (\ref{eq:JarzynskiEquality})).}
\end{figure}

Histograms of the measured work $W$ are shown in Fig. \ref{Figure3}(a) for increasing switching times $\tau$. The equivalent work follows a near-Gaussian distribution with an offset according to the second law $\left\langle W\right\rangle \geq \Delta F$ which converges to $\left\langle W\right\rangle=\frac{1}{2}\kappa_{\rm eff} \, \Delta x^2$ as the switching time increases (Fig. \ref{Figure3}(b), dashed line). For shorter times, the particle is not yet in the Soret ``equilibrium'' when switching the potential once more, so the JE is not expected to hold. The distribution of the heat $Q$ that is released while the particle relaxes considerably differs from a Gaussian distribution (see Fig. \ref{Figure3}(c)). Nevertheless, the average equivalent energy dissipated matches the average work that is done on the system for the different switching times due to conservation of energy (cf. Fig. \ref{Figure3}(b)). The exponential average over the work done $\left \langle e^{-\beta W} \right \rangle$ converges to unity for sufficiently long switching times at which the initial state is in steady state, consequently, verifying the JE for the virtual effective potential of the thermophoretic trap (see Fig. \ref{Figure3}(d)). 

As compared to  conventional methods driving a system out of equilibrium with the help of conservative external forces, our virtual effective potential is determined purely by a local temperature rise as compared to room temperature. The particle is lifted from one local temperature to a new temperature level when the target position is changed in thermophoretic trapping. This allows us to restate Equation (\ref{eq:AppliedWork}) to extract the temperature difference between the two positions via
\begin{equation}\label{eq:TempDiffW}
\Delta T_{\rm W}\left[x(t)\right]=\int_0^{\tau}\mathrm dt\, \frac{\partial \Delta T(x(t),t)}{\partial \lambda}\,\dot{\lambda}(t).
\end{equation}
The temperature change due to the displaced parabola is on average $\left \langle \Delta T_{\rm W} \right \rangle=\frac{1}{2}\alpha\Delta x^2$ and, in particular, exceeds the temperature difference of the final and initial states $\Delta T_{\rm i\rightarrow f}$ (``second law of thermophoresis''):
\begin{equation}
\left \langle \Delta T_{\rm W} \right \rangle\geq \Delta T_{\rm i\rightarrow f}.
\end{equation}
Furthermore, as it is shown in the experiment (see right axis in Fig. \ref{Figure3}), the temperature differences $\Delta T_{\rm W}$ and $\Delta T_{\rm i\rightarrow f}$ can be related via a ``thermophoretic Jarzynski equality'' 
\begin{equation}
\left \langle e^{- S_{\rm T}\Delta T_{\rm W}} \right \rangle=e^{-S_{\rm T}\Delta T_{\rm i\rightarrow f}},
\end{equation}
where $e^{-S_{\rm T}\Delta T_{\rm i\rightarrow f}}=1$ holds true in this case since $\Delta T_{\rm i\rightarrow f}$ vanishes (cf. Supplemental Material -- III. Extended Jarzynski Equality). Subsequent to each switching event, the particle cools by the value $\left \langle \Delta T_{\rm Q} \right \rangle-\Delta T_{\rm i\rightarrow f}=\left \langle \Delta T_{\rm W}\right \rangle$ (``first law of thermophoresis'') during the relaxation to its Soret equilibrium. This temperature difference is obtained from the trajectory with the relation
\begin{equation}\label{eq:TempDiffQ}
\Delta T_{\rm Q}\left[x(t)\right]=\int_0^{\tau}\mathrm dt\, \frac{\partial \Delta T(x(t),t)}{\partial x}\,\dot{x}(t).
\end{equation}
In our experiments, merely a minuscule temperature change of about $\left \langle \Delta T_{\rm W} \right \rangle=\left \langle \Delta T_{\rm Q}\right \rangle\approx 0.37\,\rm K$ drives the particle between the target positions (at $S_{\rm T}=1.0\,\rm K^{-1}$, Fig. \ref{Figure2}(d)). Such a small temperature change causes a rather long relaxation process of several hundred milliseconds as displayed in Fig. \ref{Figure3}(d). It requires more than 0.6 seconds to relax to the new Soret ``equilibrium'' of the final state. Thus, contributions of thermal non-equilibrium processes might be crucial during, for example, optical tweezing experiments as well.

\section{V. Mathematical equivalence of potential energy and temperature landscapes}
The equivalence of the irreversible transition between potential energy landscapes and temperature landscapes becomes obvious if the dynamics of the systems are modeled in the Langevin approach. Starting with the macroscopic equation of motion for a particle in an external potential superimposed by a stochastic force exerted by molecules of the surrounding \cite{book:VanKampen}, the Langevin equation reads
\begin{equation}\label{eq:GeneralLangevin}
m\ddot{x}(t) = -\gamma \dot{x}(t) + \frac{\mathrm{d}U}{\mathrm{d}x} + \sqrt{2k_{\mathrm{B}}T\gamma}\xi(t),
\end{equation}
where $\xi(t)$ is assumed to be white noise with vanishing mean, i.e., $\left \langle \xi(t) \right \rangle  = 0$, and a correlation function according to $\left \langle \xi(t)\xi(t^\prime) \right \rangle = \delta(t-t^\prime)$. In the situation that $\gamma$ and $T$ are constants, Equation (\ref{eq:GeneralLangevin}) simplifies to
\begin{align}\label{eq:SimplifiedLangevin}
\dot{x} &= \frac{1}{\gamma}\frac{\mathrm{d}U}{\mathrm{d}x} + \sqrt{2D}\xi(t) \nonumber  \\
 &= v_{\rm trap}(x,t)+\sqrt{2D}\xi(t)
\end{align}
in the overdamped regime, i.e., the inertia of the particle is negligible.

For the case that the velocity induced by the trap is the result of an external force $-\frac{\mathrm{d}U}{\mathrm{d}x}$ due to a potential energy field $U$, the motion results because of Stokes drag
\begin{equation}\label{eq:VelocityPotenital}
v_{\rm trap}=-\mu\frac{\mathrm{d}U}{\mathrm{d}x}
\end{equation}
with the mobility $\mu=1/\gamma$. The velocity for a harmonic potential is then $v_{\rm trap}=-\mu\kappa \left[x-\lambda(t)\right]$. In thermophoresis, the motion is caused by a temperature gradient along the surface of the particle according to  $v_{\rm trap}=-D_{\rm T} \frac{\mathrm{d}T}{\mathrm{d}x}$ \cite{Bregulla:PhysRevLett:2016}. With the harmonic implementation of the virtual temperature field, the thermophoretic drift velocity reads 
\begin{equation}\label{eq:VelocityTrap}
v_{\rm trap}=-D_{\rm T}\alpha \left[x-\lambda(t)\right]. 
\end{equation}
Although the microscopic origin of both effects is fundamentally different, they are described by real/effective potentials that are either the potential energy (e.g., in optical tweezing \cite{Suzuki:OptLett:1997}) or the temperature field (for thermophoretic trapping). The microscopic details of the underlying hydrodynamics are hidden in the (Stokes) mobility coefficient $\mu$ and the thermophoretic mobility $D_{\rm T}$. The mathematical description by the Langevin equation, however, is fully equivalent and related via $\kappa_{\rm eff}=S_{\rm T}\alpha \,k_{\rm B}T=k_{\rm B}T/\sigma^2$. A detailed derivation of the JE with temperature gradients starting from a more general Langevin approach with multiplicative noise that treats the inhomogeneities in the paths $\gamma$ can be found in the Supplemental Material. 

The previously described theory bases on the fact that the overall temperature increases $\Delta T(\boldsymbol{r})$ in the thermophoretic trapping setup are small compared to the ambient temperature $T_0$, although the non-isothermal trapping principle relies on strong temperature gradients in the order of $10^8~\mathrm{K\,m^{-1}}$ \cite{Braun:NanoLett:2015}. This, in turn, entails that Brownian fluctuations are not significantly altered and the probability density function of the measured equivalent work follows a Gaussian statistics. Concerning a debate on the unrestricted validity of the JE \cite{Mauzerall:JStatMechTheoryExp:2004, Jarzynski:JStatMechTheoryExp:2004, Mauzerall:MolPhys:2005}, this Gaussian statistics of the underlying noise constitutes an important prerequisite for the JE to hold. Furthermore, Cohen and Mauzerall question in \cite{Mauzerall:MolPhys:2005} the Boltzmann factor $\beta$ in the derivation of the JE to be solely dependent on the temperature of the heat bath, although the non-equilibrium paths which connect the initial equilibrium state A with the final equilibrium state B might be at different temperatures $T_\mathrm{irr} \neq T_0$. The distribution over the system's energy levels would, consequently, be different and, hence, the canonical partition function would be absent. However, our experimental conditions are adjusted such that $T_\mathrm{irr} \approx T_0$ holds.

\section{VI. Conclusion}
We have tested the Jarzynski equality for a single colloid in a thermophoretic trap by switching effective temperature profiles between two steady state positions within the trap. The confinement of the PS particle to both positions results in these experiments not from potentials, but dynamic temperature gradients being generated by a feedback loop. The JE for vanishing steady state temperature differences is verified by calculating the exponential average over the work done on the system which converges to unity as the switching times increase. Furthermore, these experiments show that already minor temperature changes of less than $1~\%$ of the room temperature are on average able to induce such relaxation processes which occur on time scales in the order of $0.1~\mathrm{s}$. The governing Langevin equation for this setup is mathematically analogous to other trapping setups suggesting that the JE is not solely valid for conservative forces. Therefore, the ``thermophoretic Jarzynski equality" might be useful to measure temperature differences in a microscopic system.

Future experiments may contribute to the debate on the general validity of the JE \cite{Mauzerall:JStatMechTheoryExp:2004, Jarzynski:JStatMechTheoryExp:2004, Mauzerall:MolPhys:2005} by strongly varying the temperature landscape in the trap which leads to strong position-dependent Brownian dynamics and, therefore, non-Gaussian statistics. Moreover, a randomized offset in the modulation of the trapping laser intensity directly perturbs the Brownian fluctuations of the system. Moreover, various statistics for the non-equilibrium distribution functions might be generated with such an approach as well exploring the necessity of the assumption of Gaussian statistics. A considerably varying temperature landscape would also shed light on the issue of the Boltzmann factor.

Finally, our thermophoretic trap also permits the experimental verification of other non-equilibrium steady state equalities as, for example, Hatano and Sasa's equality \cite{Hatano:PhysRevLett:2001, Trepagnier:ProcNatAcadSciUSA:2004} in varying temperature fields.


\section{Acknowledgement}
F.C. acknowledges financial support by the German Research Foundation (Deutsche Forschungsgemeinschaft, DFG) through the Collaborative Research Center TRR 102 `Polymers under multiple constraints: restricted and controlled molecular order and mobility' (funded by the Deutsche Forschungsgemeinschaft (DFG, German Research Foundation), project number 189853844). We thank A. Kramer for helping to revise the manuscript.
\\
%

\end{document}


\title{Jarzynski Equality for Soret Equilibria -- Virtues of Virtual Potentials\newline Supplemental Material} 
\author{Tobias Thalheim}
\affiliation{Molecular Nanophotonics Group, Peter Debye Institute for Soft Matter Physics, Leipzig University, 04103 Leipzig, Germany}
\author{Marco Braun}
\affiliation{Molecular Nanophotonics Group, Peter Debye Institute for Soft Matter Physics, Leipzig University, 04103 Leipzig, Germany}
\author{Gianmaria Falasco}
\affiliation{Theory of Condensed Matter, Institute for Theoretical Physics, Leipzig University, Postfach 100 920, 04009 Leipzig, Germany}
\affiliation{Complex Systems and Statistical Mechanics, Department of Physics and Materials Science, University of Luxembourg, L-1511, Luxembourg}
\author{Klaus Kroy}
\affiliation{Theory of Condensed Matter, Institute for Theoretical Physics, Leipzig University, Postfach 100 920, 04009 Leipzig, Germany}
\author{Frank Cichos}
\affiliation{Molecular Nanophotonics Group, Peter Debye Institute for Soft Matter Physics, Leipzig University, 04103 Leipzig, Germany}
\date{\today}

\maketitle

\newpage

\section{I. Sample Preparation}

\begin{figure}[h]
\centering
\includegraphics[width=1\textwidth]{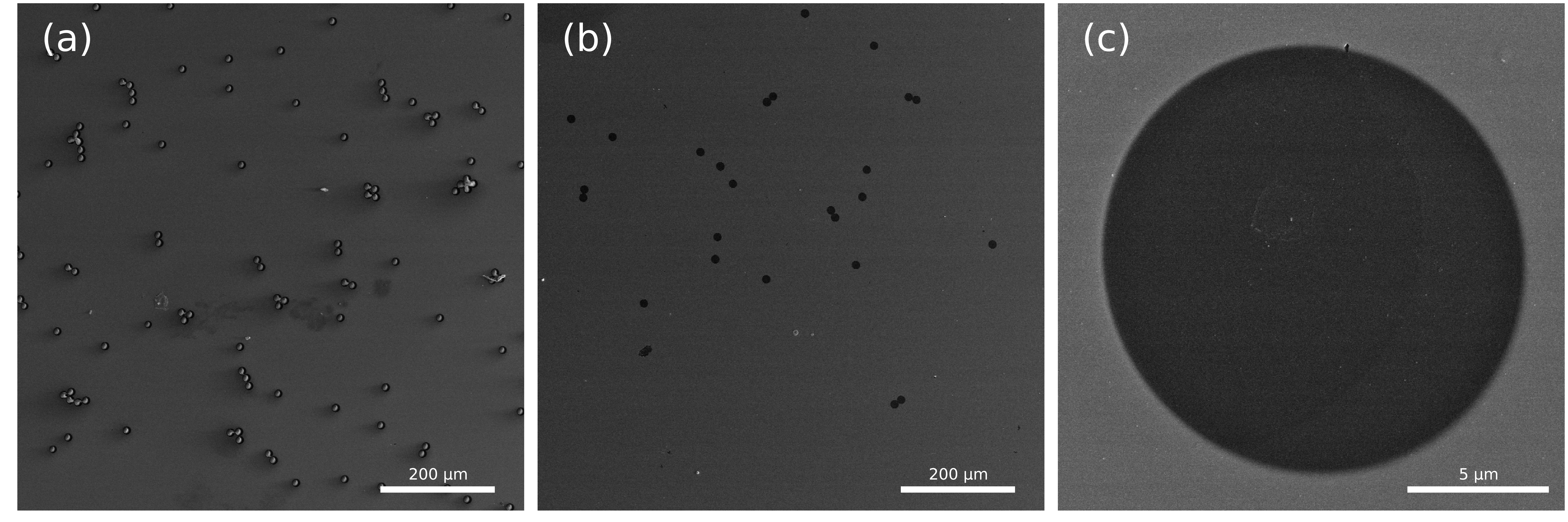}
\caption{Transmission electron microscopy images of single steps during the fabrication of thermophoretic traps. (a) PS beads spin-coated onto the chromium layer. (b) Remaining traps after ultrasonification. (c) A single trap used in the experiments.}
\label{Fig:ThermophoreticTraps}
\end{figure}

\noindent Conventional $22~\mathrm{mm} \times 22~\mathrm{mm}$ glass coverslips serve as basis for the gold traps produced by microsphere lithography. After thoroughly rinsing the coverslips with acetone, isopropyl, ethanol, Milli-Q water, and drying them with a nitrogen gun, the coverslips are additionally plasma-cleaned for $5~\mathrm{min}$. To enhance the adherence of the gold structure, the coverslips are coated with a 5 nm chromium layer by thermal evaporation. A dilute solution of polystyrene (PS) particles at a diameter of $15.2~\mathrm{\upmu m}$ (microParticles GmbH, PS-F-15.2) is spin-coated on top of the chromium layer (cf. Supplementary Fig. \ref{Fig:ThermophoreticTraps}(a)). Subsequently, the sample is covered with a 50 nm gold layer utilizing thermal evaporation once more. The thermophoretic traps remain after removing the PS beads by ultrasonification as shown in the Supplementary Fig. \ref{Fig:ThermophoreticTraps}(b). Residuals due to the PS particles or other impurities on the traps are remediated with Chromium Etch No. 1 (MicroChemicals GmbH) that removes the chromium layer which is not covered by gold, i.e., within the trapping region (Supplementary Fig. \ref{Fig:ThermophoreticTraps}(c)).

\noindent The measurement sample consists of two glass coverslips confining a thin liquid film. One of the glass coverslips contains the thermophoretic traps, whereas the other one seals the sample preventing evaporation of the solution and adjusts the height of the sample to about $500~\mathrm{nm}$. The dye-doped 200 nm PS beads (ThermoFisher, F8810) used in the experiments are, thus, not spatially confined in the trap, indeed, can diffuse freely between the two glass coverslips. The sample height is adjusted by preparing the second glass coverslip with a thin layer of conventional spray paint. The central region of the coverslip is sheltered with a two-cent coin from spray paint. Consequently, exclusively outer regions of the coverslip are covered. All coverslips are passivated with a 1 \% Pluronic\textsuperscript{\textregistered} F-127 solution (Sigma-Aldrich, P2443) to prevent sticking of the PS beads during the experiments. Approximately $0.3~\mathrm{\upmu l}$ of the diluted bead solution is pipetted between the two coverslips. A squeezing device combined with a microscope sample holder applies a homogeneous pressure to the coverslips for a proper distribution of the sample liquid between them and a reproducible height adjustment.

\section{II. Experimental Setup}
A sketched overview of the experimental setup can be found in Supplementary Fig. \ref{Fig:Setup}. The setup is built around an inverted microscope (Olympus, IX71) with a mounted piezo translation stage (Physik Instrumente, P-517.3CL). A CW laser (CNI Optoelectronics Technology Co., Ltd., MGL-FN-532-500mW) at a wavelength of $532~\mathrm{nm}$ is used to excite the fluorescently labeled PS beads. The laser is focused into the back-focal plane of the objective lens (Olympus, 100x/NA 1.3, oil) to ensure a homogeneous illumination in the center of the sample ($\omega_{0,\mathrm{wide}} \approx 20~\mathrm{\upmu m}$). The same objective lens images the fluorescence of the beads to an EMCCD camera (Andor, iXon 3) via a tube lens at a focal length of $500~\mathrm{mm}$. Laser light which is back-reflected from the objective lens is blocked with a notch filter (F in Supplementary Fig. \ref{Fig:Setup}, Thorlabs, NF 533-17) in front of the camera. The dichroic mirror (D in Supplementary Fig. \ref{Fig:Setup}) is selected according to the excitation and emission of the widefiled illumination as well as the heating laser (Pusch OptoTech GmbH, $\lambda = 532~\mathrm{nm}$). The heating laser generates the temperature gradients on the gold structure. A beam expander increases the beam of the heating laser and the objective lens focuses the heating laser into the sample ($\omega_{0,\mathrm{focused}} \approx 1.0~\mathrm{\upmu m}$).  An acousto-optic deflector (Brimrose, 2DS-90-45-532) and a lens system are used to steer the laser in the $xy$-plane of the sample. The acousto-optic deflector and the piezo translation stage are controlled with an analog-to-digital converter (ADC, ADwin-Gold \RNum{2}, J\"ager Messtechnik) which, in turn, is driven by a LabView program. The LabView program records and evaluates the fluorescence images in real time as well. A halogen lamp (Olympus, U-LH100L-3) is used to support the alignment of the trap in the camera view.

\begin{figure}[h]
\centering
\includegraphics[width=0.55\textwidth]{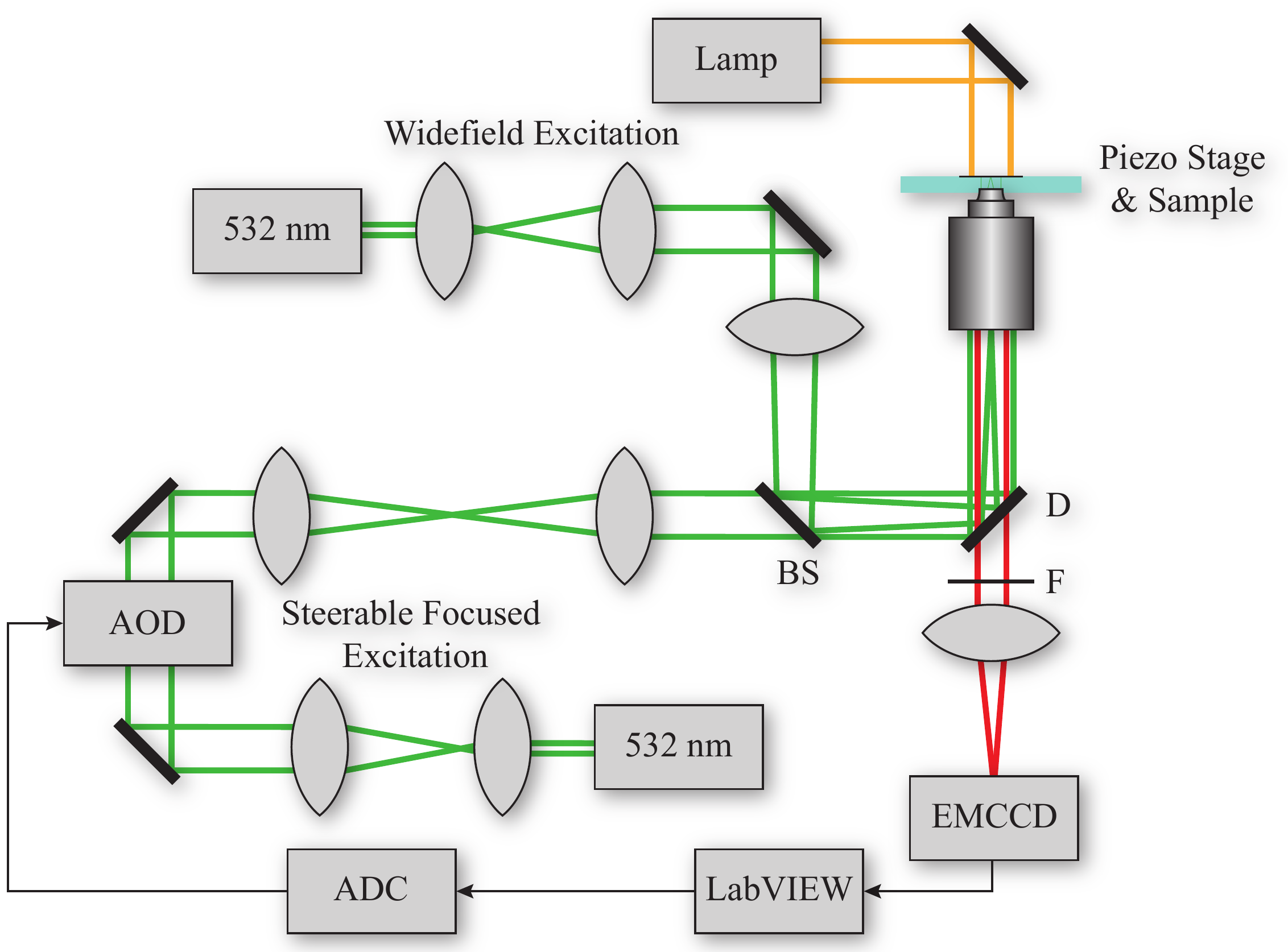}
\caption{Sketch of the experimental setup.}
\label{Fig:Setup}
\end{figure}


\section{III. Extended Jarzynski Equality}
In \cite{Jarzynski:PhysRevLett:1997}, Jarzynski derived his famous equality for a system which is in contact with a heat bath at a fixed temperature $T$. This equality -- the Jarzynski Equality (JE) -- is also valid for microscopic situations in which the ambient temperature slightly varies not altering the strength of the fluctuations of the considered Brownian particles significantly as will be shown in this Supporting Note. This situation is faced in the experiments presented in the main paper. The starting point of the derivation is Equation (26) in \cite{Kuroiwa:JPhysAMathTheor:2014}, which is a one-dimensional Langevin equation with nonlinear noise coupled with stochastic variables, i.e., a multiplicative Langevin equation:
\begin{equation}
\label{Eq:StartingPoint}
\dot{x} = -\Gamma(x)\frac{\partial U}{\partial x} + \frac{D(x)}{\Gamma(x)}\frac{\partial \Gamma(x)}{\partial x} - \alpha \frac{\partial D(x)}{\partial x} + g({x}^*)\xi (t),
\end{equation}
where $D(x)$ is the position-dependent diffusion coefficient of the Brownian particle, $\Gamma(x)$ the inverse of its  friction coefficient and $U(x)$ some potential of the system. The mass $m$ of the particle is set to one. $\xi(t)$ is white noise with
\begin{equation}
\langle \xi(t) \rangle = 0~~\mathrm{and}~~\langle \xi(t)\xi(t^{\prime}) \rangle = \delta(t-t^{\prime}).
\end{equation}
The term $g(x)$ is a position-dependent amplitude of the noise. As Equation (\ref{Eq:StartingPoint}) is meaningless as far as an interpretation of the multiplicative noise in terms of stochastic integration is missing, a more rigorous notation should be
\begin{equation}
\Delta x = f(x) \Delta t + g(x^{*})\Delta W(t)
\end{equation}
with $f(x)\coloneqq -\Gamma(x)\frac{\partial U}{\partial x} + \frac{D(x)}{\Gamma(x)}\frac{\partial \Gamma(x)}{\partial x} - \alpha \frac{\partial D(x)}{\partial x}$ and $\Delta x = x(t+\Delta t) - x(t)$. $\Delta t$ is a sufficiently small increment of time. The increment of the Wiener process $\Delta W(t)$ is given by
\begin{equation}
\Delta W(t) = \int_t^{t+\Delta t} \mathrm{d}s\xi(s)
\end{equation}
satisfying $\Delta W(t)^2 = \Delta t$ in a mean-square sense. The midpoint value $x^{*}$ is defined as:
\begin{equation}
x^{*} \coloneqq \alpha x(t+\Delta t) + (1-\alpha) x(t)
\end{equation}
for some $\alpha \in [0,1]$.

\noindent In our experimental situation, it holds that 
\begin{equation}
D(x) = \Gamma (x) T(x),
\end{equation}
where $T(x)$ is the absolute temperature varying in space (and $k_{\mathrm{B}} = 1$). Furthermore, we assume that the temperature field depends explicitly on time, i.e., it may be changed according to some known protocol $\lambda = \lambda(t)$. In our case of an instantaneous shift, a quadratic temperature field reads $T(x, \lambda(t)) \propto (x-\lambda(t))^2$ with $\lambda(t) = \hat{x}\Theta(t)$ \footnote{Also $\Gamma$, $D$ and $D_{\mathrm{T}}$ become explicit functions of time due to their dependence on $T$, but we will avoid the explicit notation for brevity.}. The potential in the first term of  $f(x)$ is the virtual effective potential introduced in Equation (6) in the main paper. For small temperature rises $\Delta T$ (as in our experiments \cite{Braun:ACSNano:2013, Parola:JPhysCondMat:2008}), the assumptions $\frac{T_0}{T(x)} \approx 1$ and $\frac{\partial S_{\mathrm{T}}(x)}{\partial x} \approx 0$ hold leading to 
\begin{equation}
-\Gamma (x) \frac{\partial U}{\partial x} = -D_{\mathrm{T}}(x)\frac{\partial T}{\partial x}
\end{equation}
in the first term of $f(x)$ which is the thermophoretic velocity. The second and third term of $f(x)$  appear in the derivation of Equation (\ref{Eq:StartingPoint}) from the corresponding inertial Langevin equation:
\begin{equation}
\ddot{x} = - \frac{1}{\Gamma(x)}\dot{x} - \frac{\partial U}{\partial x} + \frac{g(x)}{\Gamma(x)}\xi(t)
\end{equation}
(Equation (10) in \cite{Kuroiwa:JPhysAMathTheor:2014}). They simplify to 
\begin{equation}
\frac{D(x)}{\Gamma(x)}\frac{\partial \Gamma(x)}{\partial x} - \alpha \frac{\partial D(x)}{\partial x} = (1-\alpha) T(x) \frac{\partial \Gamma}{\partial x} - \alpha \Gamma (x) \frac{\partial T}{\partial x}.
\end{equation}
The systematic term $f(x)$ in our experimental system can be, consequently, reformulated to
\begin{equation}
f(x) \equiv - D_{\mathrm{T}}(x) \frac{\partial T}{\partial x} + \frac{1}{2}T(x)\frac{\partial \Gamma}{\partial x} - \frac{1}{2}\Gamma(x)\frac{\partial T}{\partial x}
\end{equation}
if we choose $\alpha = 1/2$ (Stratonovich convention). The noise amplitude $g(x)$ is in our case
\begin{equation}
g(x) \equiv \sqrt{2D(x)}.
\end{equation}

\noindent In order to obtain the fluctuation relation (and, in particular, the extended JE), the conditioned probability $P_{\lambda}(x \lvert x_\mathrm{i})$ of a trajectory starting in $x_\mathrm{i}$ at $t_\mathrm{i}$ and ending in $x_{\mathrm{f}}$ at $t_{\mathrm{f}}$ is compared to the probability of the corresponding anti-trajectory $P_{\tilde{\lambda}}(\tilde{x} \lvert x_{\mathrm{f}})$. This ansatz translates the description of the stochastic dynamics into a path integral formalism which is an alternative method to Fokker--Planck and Langevin equations. The basis of the following calculations can be found in \cite{Lau:PhysRevEStatNonlinSoftMatterPhys:2007}, cf. particularly Equation (4.13). The conditioned probability/path integral $P_{\lambda}(x \lvert x_\mathrm{i})$ is:
\begin{equation}
\label{Eq:CondPath}
\begin{split}
P_{\lambda}(x \lvert x_\mathrm{i}) &\propto \exp \Bigg(-{\int_{t_\mathrm{i}}^{t_{\mathrm{f}}}}\mathrm{d}t \Bigg[ \frac{1}{2g(x)^2} \Big( \dot{x}(t)-f(x(t)) + \frac{1}{2}g(x(t))\frac{\partial g(x(t))}{\partial x(t)}\Big)^2 + \frac{1}{2}\frac{\partial f(x(t))}{\partial x(t)}\Bigg] \Bigg) \\
& = \exp \Bigg(-{\int_{t_\mathrm{i}}^{t_{\mathrm{f}}}\mathrm{d}t \Bigg[ \frac{1}{4D} \Big( \dot{x} +  D_{\mathrm{T}}\frac{\partial T}{\partial x}} + \Gamma \frac{\partial T}{\partial x} \Big)^2 + \frac{1}{2}\frac{\partial f}{\partial x}\Bigg] \Bigg). \\
\end{split}
\end{equation}
The path integral of the anti-trajectory is obtained with the time-reversal transformations $x(t) \rightarrow \tilde{x}(t) = x(-t)$ and $\lambda(t) \rightarrow \tilde{\lambda}(t) = \lambda(-t)$:
\begin{equation}
\begin{split}
P_{\lambda}(\tilde{x} \lvert x_{\mathrm{f}}) &\propto \exp \Bigg(-{\int_{t_\mathrm{i}}^{t_{\mathrm{f}}}}\mathrm{d}t \Bigg[ \frac{1}{2g(\tilde{x})^2} \Big( \dot{\tilde{x}}(t)-f(\tilde{x}(t)) + \frac{1}{2}g(\tilde{x}(t))\frac{\partial g(\tilde{x}(t))}{\partial \tilde{x}(t)}\Big)^2 + \frac{1}{2}\frac{\partial f(\tilde{x}(t))}{\partial \tilde{x}(t)}\Bigg] \Bigg) \\
&= \exp \Bigg(-{\int_{t_\mathrm{i}}^{t_{\mathrm{f}}}}\mathrm{d}t \Bigg[ \frac{1}{2g(x)^2} \Big(- \dot{x}(t)-f(x(t)) + \frac{1}{2}g(x(t))\frac{\partial g(x(t))}{\partial x(t)}\Big)^2 + \frac{1}{2}\frac{\partial f(x(t))}{\partial x(t)}\Bigg] \Bigg) \\
&= \exp \Bigg(-{\int_{t_\mathrm{i}}^{t_{\mathrm{f}}}\mathrm{d}t \Bigg[ \frac{1}{4D} \Big( \dot{x} -  D_{\mathrm{T}}\frac{\partial T}{\partial x}} - \Gamma \frac{\partial T}{\partial x} \Big)^2 + \frac{1}{2}\frac{\partial f}{\partial x}\Bigg] \Bigg).
\end{split}
\end{equation}

\noindent The unconditioned probability $P_{\lambda}(x)$ is obtained by multiplying Equation (\ref{Eq:CondPath}) by the particle probability density $\rho(x_\mathrm{i}, t_\mathrm{i})$ evaluated at the initial time. Different choices for the particle probability density are possible which describe different processes and lead to different relations. The extended JE follows by taking for $\rho(x_\mathrm{i}, t_\mathrm{i})$ the stationary probability density at the fixed value $\lambda (t_\mathrm{i})$ of the protocol, which can, for example, be found in \cite{VanKampen:JPhysChemSolids:1988}:
\begin{equation}
\label{Eq:ProbDensity}
\begin{split}
\rho(x_\mathrm{i}, t_\mathrm{i}) &= \frac{1}{Z(t_\mathrm{i})} \exp \Bigg( -\int_{0}^{x_\mathrm{i}} \mathrm{d}x \frac{\nicefrac{\partial U(x)}{\partial x}}{T(x)}\Bigg)\\
&= \frac{1}{Z(t_\mathrm{i})} \exp \Bigg( -\int_{0}^{x_\mathrm{i}} \mathrm{d}x S_\mathrm{T}(T)\frac{\partial T}{\partial x}\Bigg).
\end{split}
\end{equation}
$Z(t_\mathrm{i})$ is the steady state partition function and $S_\mathrm{T} = \frac{D_\mathrm{T}}{D} + \frac{1}{T}$ the position-dependent Soret coefficient. The lower limit in the integration is an arbitrary constant. The $1/T$-term in the Soret coefficient matters for the case that particle--solvent interactions are absent during thermophoresis, i.e., the suspended particles are considered as an ideal gas in a temperature landscape \cite{Fayolle:PhysRevE:2008}. This term is less than $0.004 ~ \mathrm{K}^{-1}$ in our experiments and, thus, neglected in the main paper. Equation (\ref{Eq:ProbDensity}) can be further rewritten by utilizing $\mathrm{d}T(x,\lambda(t)) = \frac{\partial T}{\partial x} \mathrm{d}x + \frac{\partial T}{\partial \lambda} \mathrm{d}\lambda = \frac{\partial T}{\partial x} \mathrm{d}x$ for $t \leq t_\mathrm{i}$:
\begin{equation}
\label{Eq:ProbDensityFinal}
\rho(x_\mathrm{i}, t_\mathrm{i}) = \frac{1}{Z(t_\mathrm{i})} \exp \Bigg( -\int_{T(x=0)}^{T_\mathrm{i}} \mathrm{d}T S_\mathrm{T}(T)\Bigg),
\end{equation}
where $T_\mathrm{i} = T(x_\mathrm{i}, t_\mathrm{i})$. The particle probability density $\rho(x_\mathrm{f}, t_\mathrm{f})$ for the unconditioned probability $P_{\tilde{\lambda}}(\tilde{x})$ of the anti-trajectories reads then:
\begin{equation}
\label{Eq:ProbDensityAntiFinal}
\rho(x_\mathrm{f}, t_\mathrm{f}) = \frac{1}{Z(t_\mathrm{f})} \exp \Bigg( - \int_{T(x=0)}^{T_\mathrm{f}} \mathrm{d}T S_\mathrm{T}(T) \Bigg),
\end{equation}
where $T_\mathrm{f} = T(x_\mathrm{f}, t_\mathrm{f})$. Equation (\ref{Eq:ProbDensityAntiFinal}) is not the stationary solution corresponding to the protocol final value $\lambda(t_\mathrm{f})$, because $\mathrm{d}T \neq \frac{\partial T}{\partial x}\mathrm{d}x$ for $t \geq t_\mathrm{i}$. This choice even does not imply that the particle is in the steady state (Equation (\ref{Eq:ProbDensityFinal})) at the end of the experiment, but is only a mathematical artifice. Hence, $Z(t_\mathrm{f})$ has no physical meaning in general.

\noindent The central quantity for the derivation of the extended JE is the path integral ratio:
\begin{equation}
\label{Eq:ProbRatio}
\begin{split}
\frac{P_{\lambda} (x)}{P_{\tilde{\lambda}}(\tilde{x})} &\equiv \frac{P_{\lambda}(x\lvert x_\mathrm{i}) \rho(x_\mathrm{i}, t_\mathrm{i})}{P_{\tilde{\lambda}}(\tilde{x}\lvert x_\mathrm{f}) \rho(x_\mathrm{f}, t_\mathrm{f})} \\
&= \frac{Z(t_\mathrm{f})}{Z(t_\mathrm{i})}\exp \Bigg(- \int_{t_\mathrm{i}}^{t_\mathrm{f}}\mathrm{d}t \Bigg[\frac{D_\mathrm{T}}{D} \dot{x} \frac{\partial T}{\partial x} + \frac{\dot{x}}{T} \frac{\partial T}{\partial x} \Bigg]\Bigg) \exp \Bigg( \int_{T_\mathrm{i}}^{T_\mathrm{f}} \mathrm{d}T S_\mathrm{T}(T)\Bigg) \\
&= \exp \Bigg( \int_{t_\mathrm{i}}^{t_\mathrm{f}} \mathrm{d}t \Bigg[ S_\mathrm{T} \dot{\lambda} \frac{\partial T}{\partial\lambda}\Bigg] + \ln Z(t_\mathrm{f}) - \ln Z(t_\mathrm{i}) \Bigg),
\end{split}
\end{equation}
where we used $\mathrm{d}T(x, \lambda(t)) = \frac{\partial T}{\partial x}\mathrm{d}x + \frac{\partial T}{\partial \lambda}\dot{\lambda} \mathrm{d}t$ to rewrite the integrand as $(\dot{x}\frac{D_\mathrm{T}}{D}\frac{\partial T}{\partial x} + \frac{\dot{x}}{T}\frac{\partial T}{\partial x}) \mathrm{d}t = S_\mathrm{T} \dot{x}\frac{\partial T}{\partial x} \mathrm{d}t = S_\mathrm{T}\mathrm{d}T - S_\mathrm{T}\dot{\lambda}\frac{\partial T}{\partial \lambda}\mathrm{d}t$. As the temperature plays the role of a potential energy in the thermophoretic drift, a work rate on the colloidal particle can be defined as $\dot{W}\equiv\dot{\lambda}\frac{\partial T}{\partial \lambda}$. The extended JE is then obtained by integrating $e^{-\int_{t_\mathrm{i}}^{t_\mathrm{f}} \mathrm{d}t S_\mathrm{T} \dot{W}}$ over all possible trajectories
\begin{equation}
\label{Eq:GeneralJE}
\begin{split}
\langle e^{- \int_{t_\mathrm{i}}^{t_\mathrm{f}} \mathrm{d}t S_\mathrm{T} \dot{W}}\rangle &= \int \mathcal{D} x e^{- \int_{t_\mathrm{i}}^{t_\mathrm{f}} \mathrm{d}t S_\mathrm{T} \dot{W}} P_\lambda(x) \\
&= e^{\ln Z(t_\mathrm{f}) - \ln Z(t_\mathrm{i})} \int \mathcal{D} \tilde{x} P_{\tilde{\lambda}}(\tilde{x}) \\
&= e^{\ln Z(t_\mathrm{f}) - \ln Z(t_\mathrm{i})},
\end{split}
\end{equation}
where we used Equation (\ref{Eq:ProbRatio}) as well as $\mathcal{D}x = \mathcal{D}\tilde{x}$ and $\int \mathcal{D}\tilde{x} P_{\tilde{\lambda}}(\tilde{x}) = 1$ (the path weights are normalized to 1 \cite{Lau:PhysRevEStatNonlinSoftMatterPhys:2007}).

\noindent In the case of a mere shift of $T(x)$, $Z(t)$ does not change. That is why the right-hand side of the extended JE, i.e., Equation (\ref{Eq:GeneralJE}) becomes $1$. The work rate for this instantaneous shift at time $t_\mathrm{i}$ is $\dot{W}(t) = \Delta T_W(x_\mathrm{i}) \dot{\lambda}(t)$ with $\Delta T_W(x_\mathrm{i}) \coloneqq T(x_\mathrm{i}, t_\mathrm{f}) - T(x_\mathrm{i}, t_\mathrm{i})$. Equation (\ref{Eq:GeneralJE}), consequently, transforms into
\begin{equation}
\langle e^{- S_\mathrm{T}(T(x_\mathrm{i}, t_\mathrm{i}))\Delta T_W(x_\mathrm{i})}\rangle = 1.
\end{equation}
For minor temperature differences, all transport coefficients can be considered approximately constant. Equation (\ref{Eq:GeneralJE}) can be rewritten in this situation such that
\begin{equation}
\label{Eq:SmallTempJE}
\langle e^{- S_\mathrm{T}W}\rangle = e^{-S_\mathrm{T}\Delta F_\mathrm{ss}},
\end{equation}
where $\Delta F_\mathrm{ss} \coloneqq -\frac{1}{S_\mathrm{T}} (\ln Z(t_\mathrm{f}) - \ln Z(t_\mathrm{i}))$ is an equivalent free energy, $W \coloneqq \int_{t_\mathrm{i}}^{t_\mathrm{f}}\mathrm{d}t\dot{W}$ is a (in general) path-dependent work, and $S_\mathrm{T}$ plays the role of an equivalent inverse temperature. Equation (\ref{Eq:SmallTempJE}) holds for general protocols $\lambda(t)$. In our experimental situation with an instantaneous shift of $T(x)$, Equation (\ref{Eq:SmallTempJE}) reduces to
\begin{equation}
\langle e^{- S_\mathrm{T}\Delta T_W(x_\mathrm{i})}\rangle = 1.
\end{equation}
\\
%